\documentclass[11pt]{article}
\usepackage{amsmath,amssymb, color}
\usepackage{fullpage}

           \newcommand\om{\omega}

\newcommand{\pa}{\partial}

\newtheorem{theo}{Theorem}

\numberwithin{equation}{section}

\Large
\begin{document}
\bibliographystyle{alpha}
\title{2-D constrained Navier-Stokes equation and intermediate
asymptotics}
\author{Caglioti, Pulvirenti \footnote{Dipartimento di
Matematica, Universit\`a di Roma `La Sapienza', P.le Aldo Moro 2,
\hfill\break 00185 Roma, Italy. E-mail: {\tt caglioti@mat.uniroma1.it, pulvirenti@mat.uniroma1.it} },
Rousset\footnote{CNRS, Laboratoire Deudonn\'e,   Universit\'e de Nice, Parc Valrose, 06108 Nice
cedex 2, France.}
}
\date{}
\maketitle
\begin{abstract}
We introduce a modified version of the two-dimensional Navier-Stokes
equation, preserving energy and momentum of inertia, which is motivated by
the occurrence of different dissipation time scales and related to the 
gradient flow structure of the 2-D
Navier-Stokes equation.  The hope is to
understand intermediate asymptotics.
The analysis we present here is purely formal.  A rigorous study of this equation will be done in a forthcoming paper.
\end{abstract}

\section{Introduction}
The two-dimensional incompressible Euler equation in vorticity formulation reads
\begin{equation}
\label{Euler}
(\pa_t+u\cdot \nabla)\om(x,t)=0.
\end{equation}
Here $x\in \Bbb R^2$, $t\in \Bbb R^+$ and $u=u(x,t)\in \Bbb R^2$ is the
velocity field defined as:
\begin{equation}
\label{uomega}
u=\nabla^{\perp}\psi, \quad \psi=-\Delta^{-1}\om.
\end{equation}
Explicitely, we have :
$$
u=K*\om, \qquad K(x)=- { 1 \over 2 \pi }\nabla^{\perp}\log |x|=-\frac 1{2\pi} \frac {x^{\perp}}
{|x|^2}.$$
This equation  
 is formally  hamiltonian (we refer for example to [MTWR]). The Hamiltonian
  is the energy 
  \begin{equation}
  \label{E}
E(\om)=\frac 12\int \psi \om dx.
\end{equation}
By Noether Theorem, 
 the center of mass $M=\int x \om$ (which is related to the invariance of $E$
  with respect to the group of translations) and 
 the momentum of inertia (which is related to the invariance of $E$ with respect
to the group of rotations) $
I(\om)=\frac 12\int |x-M|^2\om dx
$
are also conserved. Moreover, due to the degeneracy
 of the Poisson bracket,   one of the main feature of \eqref{Euler}
 is the presence of an infinite number  of conserved quantities, sometimes
 called Casimir: 
 all the integrals of the form 
$$
F_{\phi}(\om)=\int \phi (\om ) dx
$$
are  conserved.  For this reason, the rigorous study of the large time
 behaviour of the solution of \eqref{Euler}
  and hence the justification from \eqref{Euler}
   of the presence of coherent structures 
observed in real and numerical experiments (see  e.g. [MS]) remains widely open.
In the following, we shall focus on non-negative solutions, 
we shall assume that 
$\om$  is  a probability distribution so that $\om$ is non negative and
$\int \om =1$ constantly in time. We also fix the reference frame in
such a way that$\int x\om =0$ for every  time. 

   One attempt to justify the appearance of these coherent structure
    is due to Onsager [O], see also [LP], [MJ]. The main idea
     is to  replace the Euler equation \eqref{Euler} by the system of
     $N$  point vortices  and to study  the Statistical Mechanics
      of these point vortices.  In a limit $N\rightarrow +\infty$, 
      the Gibbs measure associated to the point vortices concentrates to
         some special stationary solutions of the Euler equation
         (called mean field solutions), this was rigorously justified in
          [CLMP1], [CLMP2], [K] and [KJ]. These states are under the form: 
\begin{equation}
\label{MF}
\om=\frac {e^{b \psi+a \frac {|x|^2}2}} {Z}
\end{equation}
where 
$$
Z=\int e^{b \psi+a \frac {|x|^2}2}
$$
is a normalization to have $\int \om = 1$.
Recalling that $\om=-\Delta \psi$, we realize that eq.n \eqref{MF} is a
nonlinear elliptic equation. The study of this equation in connection
 with the variational principles arising from statistical mechanics
  was performed in [CLMP1], [CLMP2] .
  Let us summarize the main results.
We define the 
free-energy functional  as
\begin{equation}
F_{(b,a)}(\om)=S(\om)-bE(\om)-aI(\om)
\end{equation}
for a given pair $a<0$ and $b>0$. $F_{(b,a)}$ is defined on the space
$\Gamma$ of all the probability densities on $\Bbb R^2$ with finite
entropy, energy and moment of inertia. We define 
 the canonical variational principle as 
\begin{equation}
\label{can}
F(b,a)= \inf_{\om \in \Gamma} F_{(b,a)}(\om).
\end{equation}
Next, for $E\in \Bbb R$ and $I>0$ let us introduce the set 
\begin{equation}
\Gamma_{(E,I)}= \{\om \in \Gamma | E(\om)=E, I(\om)=I\}
\end{equation}
and  consider the microcanonical variational principle
\begin{equation}
\label{mcan}
S(E,I)= \inf_{\om \in \Gamma_{(E,I)}} S(\om).
\end{equation}
The above variational problems  can be  related to the solutions of the
Mean-Field equation \eqref{MF}. Moreover, in the whole space
 $\mathbb{R}^2$, we have also the  equivalence of the ensembles: 
\begin{theo}[CLMP1, CLMP2]
 For $a<0$ and $0<b<8\pi$:

i) There exists a unique, radially symmetric minimizer $\om=\om_{(b,a)} \in \Gamma$
of the problem \eqref{can} which  is the unique radially
symmetric solution to eq.n \eqref{MF}.

ii) When $ b\to 8\pi$, $\om$ converges (weakly) to a $\delta$ at the
origin.

iii) $F(b,a)$ is a concave smooth function and
$$
\frac {\pa F}{\pa a}= I(\om_{(b,a)}), \qquad
\frac {\pa F}{\pa b}= -E(\om_{(b,a)})
$$

iv) For $E\in \Bbb R$ and $I>0$ define
$$
S^*(I,E)=\sup _{a,b} (F(b,a)+bE+aI)
$$
and denote by $b(I,E)$ and $a(I,E)$ the unique maximizers. Then
$S(I,E)=S^*(I,E)$ and hence $S$ is a smooth convex function.

v) The variational problem \eqref{mcan} has a unique minimizer $\om (I,E)$ and
$$
\om (I,E)=\om_{(b(I,E),a(I,E))}.
$$
\end{theo}
Note that when $b\leq 0$ the theory is easier. Indeed the functional
$F_{(b,a)}(\om)$ is convex so the minimization problem is standard and eq.n
\eqref{MF} has a unique (radial) solution [GL].  We also point
 out that the equivalence between \eqref{can} and \eqref{mcan} which is established
  in v) is also useful to establish the existence of a minimizer for \eqref{mcan}.
   Indeed, in an unbounded domain,  the  existence of a minimizer
    for \eqref{mcan} seems difficult to establish directly because
     of the absence of higher moment control which would allow to pass
     to the limit in the constraint for the moment of inertia.
Note also that eq.n \eqref{MF} has a natural statistical mechanical
interpretation, its solutions being Gibbs states   with a self-consistent
interaction. Therefore $-b$ is an inverse temperature. Hence $b>0$ implies
negative temperature states, as predicted by Onsager [O] in terms of
point vortex theory.

  A rigorous justification of the fact that the solutions of the mean field equation
   plays a special part in the large time behaviour of the Euler equation
    still seems an out of reach problem.
An attempt towards the justification of the fact that the states
 \eqref{MF} play a special part in the 2D turbulence could
  come from the study of the intermediate behaviour
   of the Navier-Stokes equation. Indeed, for the Navier-Stokes
    equation,
  \begin{equation}
  \label{NS}
(\pa_t+u\cdot \nabla)\om(x,t)=\nu \Delta \om (x,t),
\end{equation}
$\int \om$ and $\int x \om$ are still conserved so that we can
 still consider non-negative solutions so that
  $\int \om =1$ and $\int x \om=0$. Nevertheless, 
due to the dissipation term in the right hand side of eq.n \eqref{NS}, the
asymptotic behavior of the solutions is trivial, namely $\om (x,t)\to 0$
pointwise and in the $L^p$ sense for $p>1$.  Consequently,
  one can hope to observe the mean field solutions only as  intermediate states.
   To formalize this idea, let us notice that  the 
momentum of inertia $I$
increases by a constant rate: 
\begin{equation}
\dot I(\om)=2\nu
\end{equation}
consequently it can be considered as constant for times $<<1/\nu$.
In a similar way, 
 $E$ and $S$  are dissipated with rates
\begin{equation}
\label{diss}
\dot E=-\nu \int \om ^2, \qquad \dot S=-\nu \int {|\nabla \om
|^2 \over \om}.
\end{equation}
Looking at eq.ns \eqref{diss},  one realizes that the energy 
 could also  evolve  on a different and longer scale of times with
respect to
$S$ (whenever the last term in \eqref{diss} dominates on the first one). 
This would suggest to consider, 
in the first
approximation, $E$ and $I$ as constant, by looking at a master equation
which modifies the Navier-Stokes equation leaving constant both energy
and moment of inertia, but retaining all the other features of the
Navier-Stokes dynamics. The derivation of such an equation, based on
 geometric arguments
 is the aim of the following section.
 
 \section{Derivation of the model}
 A natural and fruitful way to approach the problem is to invoke a recent
characterization of the Navier-Stokes equation connected with
  the mass
transport problem and the associated differential calculus introduced
 in \cite{Otto}. We follow the excellent
monographies \cite{V}, \cite{AGS} for outlining the main ideas.

Let $\cal{M}$ be the manifold of the probability measures on $\Bbb R^2$.
One can formally give to $\mathcal{M}$ a structure of Riemannian manifold.  
For any $\rho \in \cal {M}$ we parametrize  the tangent space as
\begin{equation}
\label{tangent}
T_{\rho} \mathcal{M}=\{\dot \rho |\dot \rho =- \hbox {div} \,u \}.
\end{equation}
 Eq.n \eqref{tangent} expresses the tangent
space to any point $\rho$ of
$\cal{M}$  as  mass preserving  velocity  vectors $\dot \rho$. 
Next, we define a Riemannian metric. On the tangent space  
$T_{\rho}\mathcal{M}$,   we define 
 a scalar product   by
\begin{equation}
\label{sp}
\langle \dot \rho_1, \dot \rho_2 \rangle_W = \int \rho^{-1} u_1 \cdot
u_2 \, dx,
\end{equation}
being $\dot \rho_i= - \hbox {div} \, u_i$, $i=1,2$. 
The gradient
$\nabla_W$ with respect to  this Riemannian   metric  
of  a  functional $F:\cal{M}\to \Bbb R$, is defined  as:
$$
\langle \nabla_W F, \dot \rho \rangle_W = DF\cdot \dot \rho= -\int 
\frac {\delta F}{\delta \rho} \hbox {div}  \,u,
$$
where $DF$ is the differential of the map $F$ and $\delta F/\delta \rho$
 the usual variational derivative. An explicit computation
shows that
\begin{equation}
\label{gradient}
\nabla_W F= - \hbox {div}\bigl[\rho \nabla
\frac {\delta F}{\delta \rho}\bigr].
\end{equation}
In a similar way, we  can define  on  $T_{\rho} \mathcal{M}$
 a skew-symmetric operator $J_{W}$ by
\begin{equation}
\label{defJ}
J_{W} \dot{\rho} =  - \mbox{div } \Big(  u^\perp  \Big), \quad
 \dot{\rho}= - \mbox{div }u
 \end{equation}
where $u^{\perp}=(u_{2},- u_{1})$. 
Note that  this yields in particular the expression
\begin{equation}
\label{gradperp}
J_{W}\nabla_W F= - \hbox {div} \big [\rho \nabla^{\perp}
\frac {\delta F}{\delta \rho} \big ].
\end{equation}

Gradient flows with respect to  a functional $F$ 
 are  the solutions to
\begin{equation}
\label{flowgrad}
\pa_t \rho = - \nabla_W F.
\end{equation}
 and    are dissipative in the sense that
$$
\frac d {dt} F= - \int  \omega  |\nabla
\frac {\delta F}{\delta \rho}|^2\leq 0
$$
whereas 
 hamiltonian flows defined by  
 $$ \partial_{t} \rho = J_{W} \nabla_{W} F $$
  are conservative since
$$ \frac d {dt} F=  0.$$

Since
 $\delta E / \delta \omega = \psi$,  we get from \eqref{gradperp}
 that  
$$
\hbox {div}\Bigl[ \om\nabla^\perp  \psi\Bigr]=J_{W} \nabla_{W} E
$$
and hence, the Euler equation can be interpreted as an Hamiltonian
 flow in this framework.

Next, to interpret the dissipative part of the Navier-Stokes equation,  we notice that
$$  \Delta \omega  =  \mbox{div }\Big( \omega \nabla {\delta S \over \delta \omega }\Big) = - \nabla_{W} S$$
where $S$ is the entropy functional.
 
 In conclusion the
Navier-Stokes equation can be expressed in terms of a gradient  and
an antigradient flow : 
\begin{equation}
\label{nsgrad}
\pa_t \om =- \nu \nabla_W S+J_{W} \nabla_W E.
\end{equation}

According to the previous discussion we assume  that the energy and the 
moment  of inertia are varying much more slowly than the entropy
functional. Therefore it may be useful to derive an effective equation
according to the following prescription. We  shall take the orthogonal
 projection (with respect to the scalar product \eqref{sp}) of the
  vector field in the  right hand side
of eq.n  \eqref{nsgrad} on the manifold  $E=$ const and $I=$ const, with the aim to
characterize the states which are close to the true dynamics 
for a large interval of times (coherent structures), as asymptotic
states of the new dynamics. We shall see that such states are the
solutions to the mean-field equation \eqref{MF}.
Since:
$$
\nabla_W E(\om)=-\hbox {div}\bigl[\om \nabla \frac {\delta
E(\om)}{\delta \om}\bigr] = -\hbox {div} \bigl[\om \nabla\psi\bigr],
$$
$$
\nabla_W I(\om)=-\hbox {div}\bigl[\om \nabla \frac {\delta
I(\om)}{\delta \om}\bigr] =-\hbox {div}\bigl[\om \nabla \frac
{x^2}{2}\bigr]= -\hbox {div} \bigl[\om x\bigr].
$$
we find  that
$$
\langle \nabla_W E,  J_{W}\nabla_W E\rangle_{W} =0
 \qquad \langle \nabla_W I \cdot  J_{W}\nabla_W E\rangle_{W}=0,
$$
and hence $J_{W}\nabla_W E$ is tangent to the manifold $E=$const, $I=$const.
On the other hand the projection of $\nabla_W S$ on the tangent space of such a manifold is
of the form 
$$ \nabla_{W} S -b\phantom{,} \hbox {div} (\om \nabla\psi)-a\phantom{,}\hbox {div} (\om
x), $$ with $a$ and $b$ two suitable multiplicators. As a
consequence,  the equation we are looking for is:
\begin{eqnarray}
\label{nscons}
\pa_t \om +u\cdot \nabla \om & = &\nu \,  \hbox {div}(\nabla \om -
b \om \nabla\psi-a \om x)   \\
\nonumber & =&  \nu\,  \hbox {div}\bigl[\om \nabla (\log \om -
b\psi-a \frac {x^2}2)\bigr],
\end{eqnarray}
with $a$ and $b$  to be determined by the simultaneous conservation of
$E$ and
$I$. 
A straightforward computation yields:
\begin{equation}
\label{ba}
b=\frac {2I \int \om^2 +2V}{2I \int \om |\nabla\psi|^2 -V^2},  \quad
a=-\frac {2 \int \om |\nabla\psi|^2+V \int \om^2}{2I \int \om
|\nabla\psi|^2 -V^2},
\end{equation}
where
\begin{equation}
\label{V}
V=\int \om x \cdot \nabla \psi=\int dx \int dy \om (x) \om (y)
x \cdot \nabla g(x-y)=-\frac 1 {4\pi}.
\end{equation}
Note that we have by the Cauchy-Schwarz inequality
\begin{equation}
\label{CSV}
V^2=(\int \om x \cdot \nabla \psi)^2 \leq 2I \int \om |\nabla\psi|^2 , 
\end{equation}
and hence $b $ is positive if inequality \eqref{CSV} holds strictly and $\int
\om^2 >\frac 1 {4\pi I}$.

We point out that eq.n \eqref{nscons} has been derived in [Ch2]  (see also [Ch1]),
by using different argoments than thse of the present paper.

  Note also that  another  class
 of similar equations was introduced in [RS] and studied mathematically
  in [MR]. Nevertheless, despite to  some formal analogy the mathematical
   properties of the equations studied in [MR] are very different
    from the ones presented here.

Let us now discuss what we may expect about  the asymptotic behavior of eq.n \eqref{nscons}. 
The main feature of eq.n \eqref{nscons} is the decay  of the entropy
functional. Indeed, we have 
\begin{equation}
\nonumber
\frac {d S(\om)}{dt}= \frac {d S(\om)}{dt}-b\frac {d E(\om)}{dt}-a \frac
{d I(\om)}{dt}=
\end{equation}
$$
\nu \int (\frac {\delta S}{\delta \om}-b \frac {\delta E}{\delta \om}
-a \frac {\delta I}{\delta \om})
\hbox {div}\bigl[\om \nabla (
\frac {\delta S}{\delta \om}-a \frac {\delta I}{\delta \om}-b
\frac {\delta E}{\delta \om})\bigr]=
$$
$$
-\nu \int \om 
\big|\nabla (
\frac {\delta S}{\delta \om}-a \frac {\delta I}{\delta \om}-b
\frac {\delta E}{\delta \om})\big|^2=
$$
$$
-\nu \int \om \big| \nabla (\log \om -b \psi -a \frac {|x|^2}2\big|^2.
$$
In particular, this suggests formally  that 
 the asymptotic 
 states satisfy
$$
\om \nabla (\log \om -
b\psi-a \frac {|x|^2}2)=0
$$
and hence the mean field equation \eqref{MF}.

We remark that 
  the procedure of constructing dissipative equations
leaving invariant a given quantity is not unique. For instance the heat
equation in $\Bbb R^2$: 
\begin{equation}
\nonumber
\pa_t \om =\Delta \om
\end{equation}
can be modified to leave invariant $I$ according to the procedure
suggested by the gradient flow structure. The result is
\begin{equation}
\label{FP}
\pa_t \om=\hbox {div}(\nabla \om - a \om x),
\end{equation}
where
$$
a=-\frac 1I.
$$
On the other hand we could also have
\begin{equation}
\label{circ}
\pa_t \om=\pa^2_{\theta,\theta}\om.
\end{equation}
Note that eq.ns \eqref{FP} and \eqref{circ} have different asymptotic states.

An attempt to characterize an intermediate asymptotics was presented  by Gallay and Wayne \cite{GW}  according to the following ideas. 

It is well known that a special solution to eq.n \eqref{NS} (for $\nu=1$) is
given by the so called Oseen vortex:
$$
 \omega (x,t)=\frac 1 {4 \pi (t+1)} e^{-\frac {|x|^2}{4(t+1)}}.
$$
 Note that this is also a solution to the heat equation. 
 It was shown  in  \cite{GW} that this solution describes the long time asymptotic of the Navier-Stokes  equation in $L^1$.
 Indeed, with the  change of variables
$$
 \xi =  \frac {x}{\sqrt {1+t}}; \qquad \tau =   \log (1+t),  \qquad 
 \om (x,t)=(1+t)^{-1} \, w (\xi, \tau),
$$
the Navier-Stokes equation in the new variables is under the form :
\begin{equation}
\label{nsresc}
\partial_{\tau} w  + v \cdot \nabla_{\xi} w = 
 \Delta_{\xi} w +\,\nabla_{\xi} \cdot \Bigl(\frac 12 \xi w \Bigr). 
 \end{equation}
 It is possible to show that $w \to W$ 
 in $L^1$ as $t \to \infty$,
where $W(\xi)$ 
is the rescaled Oseen vortex. As a consequence the Oseen vortex can be
thought as characterizing an intermediate asymptotics before the
dissipation scale. Note that  $W$ is also a solution to \eqref {MF} for $b=0$.

This analysis enters perfectly in the context of
the projected gradient flows. Indeed neglecting the energy, the
mere constance of $I$  yields
\begin{equation}
\partial_{t} \omega  + u \cdot \nabla \omega = 
 \Delta \omega + \frac 1I \nabla \cdot\Bigl(\omega x) \Bigr),  
\end{equation}
that is eq.n \eqref{nsresc} for $I=2$.  In this particular case we have seen how eq.n
\eqref{nsresc} can be obtained also by a simple change of variables, due to the
fact that the dissipation rate of $I$ is constant. Roughly speaking the analysis we present here is an attempt to see what happens before the appearence of the Oseen vortex  by projecting on a less robust manifold.
Indeed one could argue that $I$ is more stable than $E$ in
many interesting physical situations. If so eq.n \eqref{nscons} should be more
appropriate on the time scale when $E$ is practically constant, while eq.n
\eqref{nsresc} should describe the fluid when $E$ start to be dissipated at
constant $I$. After that, everything disappears.

It would be interesting to look for a numerical evidence of this fact, if
true.

The formal derivation of eq.n \eqref{nscons} is also in agreement with the stochastic vortex theory as we are going to illustrate.

As explained previously, the procedure of constructing dissipative equation
leaving invariant a given quantity is not unique. Equation
 \eqref{nscons} has been interpreted  as a constrained Navier-Stokes flow
 and it turns out that  the asymptotic states are solution of
 the microcanonical variational principle. Since this variational
 principle is obtained by Mean Field limit  from the statistical 
  theory of the point vortices, it is interesting to see how equation \eqref{nscons}
  is related to the theory of point vortices.  
Now we show how, at least formally, the structure of the
constrained gradient flow is compatible with the discretization of the
Navier-Stokes equation obtained by mean of the stochastic vortex theory.
Since in this context we are not interested in the asymptotic behavior,
we limit ourselves in considering the energy constraint only. Also the 
viscosity does not play any role so we set $\nu=1.$ 

Consider  
$N$ stochastic vortices in
$\Bbb R^2$. They obey the stochastic differential eq.n:
\begin{equation}
\label{vortices}
 dx_i = \frac 1 N \sum _j  \nabla ^{\perp} g (x_i-x_j) dt +\sqrt {2}dw_i
\end{equation}
where $ \{w_i\}_{i=1}^N $ are $N$ independent standard Brownian motions.
Here $g(x)$ is a regularization of the Green fuction $-\frac 1{2\pi}\log
|x|$.  It is well known (\cite{MP2}, \cite{Os}...) that the empirical random measure
\begin{equation}
\mu_N (dx,t) = \frac 1 N \sum _{j=1}^N   \delta (x_j(t)-x) dx
\end{equation}
approaches the solution to the (regularized) Navier-Stokes equation, if it
happens at time zero.

We now consider the  mean-field energy:
$$
H(x_1 \dots x_N)=\frac 1 N \sum _{j<r}   g (x_j-x_r).
$$
In order to guarantee the condition
$$
dH=0
$$
a short computation using the Ito formula shows that we have to modify the
process according to:
\begin{equation}
\label{vortexP}
 dx_i = \frac 1 N \sum _j  \bigl[\nabla ^{\perp} g (x_i-x_j)-
 b_N (t) \nabla g (x_i-x_j)\bigr] dt +\sqrt {2}dw_i+
\end{equation}
$$
\sum_{i,j} D^2_{i,j} H \frac { \nabla_i H \cdot \nabla_j H} 
{|\nabla H |^4 } dt
- \frac 1{|\nabla H |^2 } \sum _j \nabla_j H \cdot dw_j \nabla_i H ,
$$
where
$$
b_N (t)= \frac {\Delta H}{ |\nabla H |^2}=\frac {\int d\mu_N \Delta g
* \mu_N}{ \int d\mu_N |\nabla g * \mu_N |^2 }.
$$
Note that the above expression makes sense only if we regularize $g$ and
this explains why we did it.

We observe that an analysis on the size of the
last two terms in \eqref{vortexP} shows that they should be negligible in the
limit
$N\to\infty$. Thus we introduce the essential process defined by
$$
 dy_i = \frac 1 N \sum _j  \bigl[\nabla ^{\perp} g (y_i-y_j)-
b_N (t) \nabla g (y_i-y_j)\bigr] dt +\sqrt {2}dw_i.
$$
Now if we replace $\mu_N (dx,t) $ by $\om (x,t)dx$ in the limit
$N\to\infty$, each process $x_j$ approaches  the nonlinear (in the McKean
sense, see \cite{Mc},\cite{MP2},\cite{Os}) process solution of
\begin{equation}
\label{esproc}
 dy =  \bigl[\nabla ^{\perp} g *\om  (y)-
b(t) \nabla g *\om(y)\bigr] dt +\sqrt {2}dw.
\end{equation}
where
$$
b(t)= \frac {\int dx \,\om \Delta g
* \om}{ \int dx \, \om |\nabla g * \om |^2 }.
$$
From eq.n \eqref{esproc},  we also derive the backward Kolmogorov equation for the
probability distribution $\om$ of the proces $y$:
\begin{equation}
\label{nsconsg}
\pa_t \om + \nabla \cdot  (u \omega)=\nabla \cdot (\nabla \om -
b \, \om \nabla g * \om).
\end{equation}
where $u=\nabla ^{\perp} g *\om $.
It is remarkable that such an equation is the microcanonical equation
\eqref{nscons}, for $a=0$, if we replace  $g$ by the true Green
function in eq.n \eqref{nsconsg},

\

\end{document}